\documentclass[12pt]{article}
\begin{document}
\newcommand{\beq}{\begin{equation}}
\newcommand{\eeq}{\end{equation}}

\title{Enhanced cohesion of matter on a cylindrical surface}
\maketitle
\author{ M.K. Kostov, M.W. Cole and G.D. Mahan \\{\em Department of Physics, The Pennsylvania State University,\\ University Park, PA 16802} }

\author{ C. Carraro \\{\em Department of Chemical Engineering,\\ University of California, Berkeley, CA 94720}}

\author{ M.L. Glasser \\{\em Department of Physics and Center for Quantum Device Technology,\\ Clarkson University, Potsdam, NY 13699}}


\begin{abstract}

We evaluate the cohesive energies E$_b$ of four systems in which particles move on
a cylindrical surface, at fixed distance R from the axis. We find quite
nonuniversal dependences of E$_b$ on R. For the Coulomb binding problem, E$_b$ is a
monotonically decreasing function of R. For three problems involving
Lennard-Jones interactions, the behavior is nonmonotonic; E$_b$ is larger at
R = $\infty$ than at R=0; the maximum binding corresponds to R $\sim \, 0.7 \, \sigma$ (the
hard core parameter). Consequences of the enhanced binding are discussed.

\end{abstract}


The discovery of carbon nanotubes has stimulated a rapid evolution of
ideas, experiments and understanding concerning states of matter
confined to the proximity of a cylindrical surface\cite{sin,mer,dre}. Examples of such
systems include electrons present within a nanotube and atoms or
molecules moving just outside or within such tubes. This paper reports
unexpected  behavior we have found in studies of four such
systems: a $+/-$ pair of charges bound by the Coulomb interaction, a pair
of atoms interacting with a Lennard-Jones (LJ) interaction, an ensemble
of $^{4}$He atoms, which condenses, and a low density fluid consisting of classical atoms. We assume that all particles move on a cylindrical surface, of radius R and
infinite length. The  assumption of surface confinement simplifies the calculations without sacrificing the basic physics.

For each of these four systems, considerable attention has been
directed previously to the investigation of two extreme limits of the present
problem. The limit R = $\infty$, here called "flatland", is that of particles
moving on a plane, i.e., a two-dimensional ($2$d) problem. This has been
extensively pursued in connection with both the $2$d electron gas and
monolayer films \cite{bru,qhe,two}. The opposite limit, R approaching zero, is here called
"lineland", a $1$d limit. Matter in lineland has been explored for many years as an abstract problem 
\cite{tak}  and has recently received particular attention in connection with the possible realization
of $1$d phases within interstitial channels or external surface grooves
on nanotube bundles 
\cite{mer,bor,kro,bon1,bon2,car,mig,hal,kar,col}.
 A logical question addressed in this paper is
whether the properties of matter in "cylinderland" evolves smoothly (or
even monotonically) between these limits as the value of R is varied. We find that the answer is ``yes'' in just one of the four cases and that intriguing  behavior arises in all four cases.

The first problem we address is the ground
state binding energy E$_b$(R) of the Coulomb interaction problem. Consider a charge $+$ e fixed in
position (at z=$\phi$=$0$ on the cylindrical surface) and a negative
charge $-$ e, mass m, which is free to move on the cylinder in the
presence of an interaction $- e^2 / r$, where r is the charges' separation; the electron's position is at cylindrical coordinates (z, $\phi$) and 
$\displaystyle{ r = \sqrt{z^2 + [2R\sin(\phi/2)]^2}}$. The Schr$\ddot{o}$dinger equation for the hydrogen atom \, is exactly solvable in $2$d,
resulting in a ground state wave function  $\displaystyle{\exp(-2r/a_0)}$, with binding energy  $\displaystyle{E_b(\infty)=4}$ Rydbergs
and mean separation $\displaystyle{ a_0 / 2}$ between the interacting charges. Here, a$_0$ is the Bohr radius.  In the opposite limit of lineland, we encounter a well-known logarithmic
divergence of E$_b$ as R approaches zero; the particle "falls to the
center of the attractive force" in $1$d. We expected the behavior of the
cylindrical problem to shift between these limits when R becomes of
order a$_0$. Figure $1$ depicts the result of a three-parameter variational
solution of the problem; details \, of the trial wave function will be
provided in a future publication (as will those of the other calculations described here). One observes in Fig. $1$ that E$_b$ increases monotonically as the reduced curvature $C = a_0/R$ is increased; the known
$1$d and $2$d limits are accurately reproduced by this calculation.  E$_b$ is not significantly affected by the nonzero curvature for $C < 1$; its value is only $\sim 1\%$ higher at $C=1$ than at $C=0$. This weak dependence becomes comprehensible when it is compared with the result from a perturbation theory, described elsewhere, where the curvature $C$ is the perturbation. The resulting (lowest order) dependence on $C$ is of the form $\displaystyle{ E_b(C) - E_b(0) \approx (1/128) C^2}$. As seen in the inset to Fig. $1$, this function is consistent with the variational result, up to $C=1$, explaining the initially weak dependence on $C$. 

The second problem we consider is the ground state cohesive
energy (per atom) of a fluid consisting of $^4$He atoms whose nuclei are confined to the
cylindrical surface. The analogous problem has been extensively investigated in $2$d \cite{whi}, for
which it is relevant to superfluid films, and $1$d, where it has been
studied for potential application to an interacting interstitial fluid\cite{bor,kro,bon1,bon2}. The cylinderland problem of $^4$He has received some attention in connection with endohedral
adsorption within nanotubes\cite{gat}. In $3$d, the cohesive energy of $^4$He is E$_b=7.2$ K. In $2$d, it is $0.85$ K and in $1$d it is $\sim 3$ mK. What R dependence is expected for E$_b$  on a cylinder?

We have studied this problem variationally, using a Jastrow trial wave function, i.e. a symmetrized product of two-particle functions
that prevent hard-core overlap of the atoms. The interatomic potential assumed in the
calculation is the modern "Aziz" potential \cite{azi}. The results of this liquid state
calculation appear in Figure $2$. The dependence of E$_b$ on
curvature is $\it {not}$ monotonic. Indeed, the binding energy is a factor $3.7$
higher near R=$1.8$ \,\AA \,  than in flatland. The nonmonotonic dependence on $C$ and lowest value at $C=\infty$ stand in stark contrast to the monotonically $\it {increasing}$ variation with $C$ found for the Coulomb problem.

The origin of the maximum binding energy near R=$2$ \,\AA \,  is  the minimum in 
the He-He interaction  near r$_{min}=2.9$ \,\AA \, . The He fluid has a strong binding
if the geometry encourages the particles to have such a spacing. This
is the case for the cylindrical geometry, as indicated by the
following argument. We introduce a function called the specific area
function $a(r)$, defined as the area on the cylinder's surface at
distance r from a specified point on the surface, per unit distance
from this point. Letting this point be the origin, we have 
$a( r) = \int d^2 r^{\prime} \,  \delta \left(\left| r^{\prime} -  r \right| \right)
$. In the flatland limit, $a(r)= 2\pi r$, while  in the
lineland limit (R$\rightarrow 0$), $a= 2\pi$R. In the cylindrical case, $a(r)$ is proportional to an elliptic integral, which exhibits a logarithmic
divergence at r = 2R. The origin of this divergence  is that a particle on one side of the cylinder has a divergent specific area at distance 2R; the sphere of this radius is tangent to the cylinder. As a consequence, the system's energy is lowered when this distance is such that the interaction is strongly attractive. The optimal binding does not occur precisely when the minimum in the He-He potential coincides with the diameter of the cylinder, but instead  at $\sim 20 \%$ higher value of R. The difference arises from the zero-point energy of the system, which expands the nearest neighbor distance beyond r$_{min}$. The same behavior occurs  in $2$d and $3$d  $^4$He; the ``nearest-neighbor'' peak in the radial distribution function occurs at distance about\, $20 \%$ greater than r$_{min}$ \cite{hall}.

What consequences accompany the enhanced binding at this value of R? Typically (but not always), strongly cohesive systems exhibit a relatively large speed of sound. Here, the sound propagation speed $s$ is derived from the relation appropriate to longitudinal density fluctuations propagating parallel to the axis of the cylinder: $\displaystyle{M s^2 = \rho^2 \frac{d^2(E/N)}{d \rho^2}}$\, . Here, M is the atomic mass and $\rho$ is the $1$d density; the derivative is evaluated at the ground state density. According to our calculations, $s= 230$ m/s at the equilibrium density $\rho = 0.3$\, \AA$^{-1}$ \, for the optimally binding radius, R = $1.8$\, \AA \, . This value may be compared to the values $s= 240, 90$ and $8.0 (3.0)$ m/s in $3$d, $2$d and $1$d, respectively. We observe that the speed  in cylinderland is $\it {significantly}$ enhanced relative to both the $1$d and $2$d values. 

Since the cylindrical fluid is a $1$d system, from the perspective of statistical mechanics, it undergoes no phase transitions at finite T. There is, however, a T=$0$ transition as the system evolves from a liquid-vapor coexisting ground state to a disordered fluid at nonzero T, with a singular heat capacity (proportional to $\delta$(T)). The integrated specific heat is a monotonic function of the binding energy. This might be observable, due to inhomogeneity in the system, as a smeared out maximum in the low T specific heat. The fluid's compressibility diverges as $\exp[E_b / (k_B T)]$ at low T, which should be observable in an adsorption isotherm. However, we have thus far no definite calculations to compare with experiments.

The third problem we address is the curvature-dependent binding energy E$_b$(R) of an atomic dimer on a cylinder of radius R. This is the two-body version of the liquid helium problem just discussed, except that here we treat the general case of atoms interacting with an arbitrary Lennard-Jones (LJ) interaction. By scaling distances relative to the hard-core diameter $\sigma$ and the energy relative to the well depth $\epsilon \, \displaystyle{\left(\tilde{z}=\frac{z}{\sigma}\, ; \tilde{R}=\frac{R}{\sigma}\, ; \tilde{r}=\frac{r}{\sigma}\, ;\,  \mathcal{E}_b = \frac{E_b}{\epsilon}\right)}$,  the Schr$\ddot{o}$dinger equation becomes:

\beq
- \eta \left(\frac{\partial^2}{\partial \tilde{z}^2} + \frac{1}{\tilde{R}^2}\frac{\partial^2}{\partial \phi^2}\right) \Psi_0  + \left[4(\tilde{r}^{-12} - \tilde{r}^{-6}) - \mathcal{E}_b \right] \Psi_0 = 0  \, .
\eeq

Note that the ground state solution to this equation is determined by the boundary conditions and the value of the dimensionless  $\it {de \,   Boer \,  quantum \,  parameter}$ $\displaystyle{\eta = \frac{\hbar^2}{M \epsilon \sigma^2}}$\,.  A very large value of $\eta$ implies a large zero-point energy and absence of any bound state. Here, we focus on a specific question: what is the threshold value ($\eta = \eta_t$) separating those problems for which the dimer exists ($\eta < \eta_t$) from those for which it does not exist? The corresponding threshold value of $\eta_t$ is known for the limiting cases of $1$d ($\eta_t = 0.1788$)\cite{kro} and $2$d ($\eta_t = 0.269$)\cite{mil}. We have answered this question for cylinderland by computing the ground state energy variationally, identifying $\eta_t$ from the point when the attractive potential energy becomes too weak for the dimer to be bound. Because the calculation is variational, the computed threshold $\eta_t$ is a lower limit to the exact value. Our $\eta_t$ results agree well with the known $1$d and $2$d limits cited above. For the general cylindrical case, the threshold value is shown in Fig. $3$. Qualitatively, it exhibits the same phenomenon as was seen in the $^4$He liquid binding problem. That is,  the binding is particularly large when the diameter of the cylinder is such that the interatomic interaction across the cylinder is strongly attractive. There is a small difference between the ``optimal'' radius, R$_{opt}$, values for the two problems: R$_{opt}/ \sigma$ has the value $0.7$ for the dimer problem and $0.77$  for the $^4$He binding problem.

Having in mind the interesting behavior of the threshold $\eta_t$ in the cylinderland, we next address the problem of the  binding energy(E$_b$) of $^4$He and $^3$He dimers on a cylinder. Employing the same variational approach as for $\eta_t$, we compute the ground state energy for these systems as a function of $\sigma / R $ ( Fig. $4$). Not surprisingly, we find significantly enhanced binding energy for the   $^4$He and $^3$He dimers at a particular range of values of R($\sim 0.65 \, \sigma$). The enhanced binding in the cylindrical geometry is particularly dramatic for the $^3$He dimer, yielding an increase in E$_b$ of $7$ orders in magnitude compared to the $2$d limit for E$_b$.

Having established that the $^3$He dimer is so  strongly bound for radius R $\sim 0.65 \, \sigma$, we now discuss the ground state of the system of many $^3$He atoms. No condensed liquid exists in either $2$d or $1$d, but perhaps one exists in cylinderland. The existence of the stable dimer does not ensure the existence of a stable N-mer for any N$\, > 2$ (as was shown explicitly in the $2$d case\cite{bru2}). One scenario is that the
ground state of the system is a gas of such dimers (analogous to H$_2$ at room
temperature). A second possibility is that the dimers coalesce to form a liquid, analogous to liquid H$_2$ between its triple point and  its critical point. A third scenario is that the dimers dissolve into a many-body liquid ground state, analogous to liquid $^3$He in $3$d. None of the first two of these possibilities has been considered previously; the question of which phase, among the three candidates, is the actual ground state remains open.

Finally, we address a fourth problem concerned with matter in cylinderland $-$ the second virial coefficient of a classical gas. By analogy with well known problems in other dimensions, we  write a low density (high T) expansion of the $1$d pressure $\displaystyle{P = \rho^2 \left(\frac{\partial f}{\partial \rho}\right)}$ , where $f$ is the free energy per particle, as $\displaystyle{P / (\rho k_B T ) \approx 1 + \frac{\rho B(T)}{2 \pi R}  + ... }$ ,  with 

\beq
B(T) = \frac{1}{2} \int d^2 {\bf r }\left[1 - \exp[- \beta V(r)] \right] \,  .
\eeq

Here, $\beta = 1 /(k_B T)$ and the integration is over the cylinder's surface. We assume the usual LJ form of interaction.
The results(to be reported in detail elsewhere) exhibit similar qualitative behavior to that found for the preceding two problems, both of which involve an interaction with a well. As in the familiar $2$d and $3$d contexts, $B$  is positive at high T due to the repulsive interactions and $B < 0$ at low T, where attractive effects dominate. The Boyle temperature T$_B$ is that for which $B$ vanishes, meaning that (within this expansion) P for the interacting system is the same as that of a noninteracting gas at the same $\rho$ and T. The results in Fig. $5$ indicate that the Boyle temperature is the highest for a cylinder of radius $R= 0.7 \, \sigma$. In the van der Waals theory of condensation, the critical temperature is $8$T$_B$/$27$.  In that mean field theory, therefore, the critical temperature is highest for a cylinder of this ``optimal'' radius. While this transition does not occur in the exact theory, one expects the virial expansion to apply at low density and here the data of Fig. $5$ imply an onset of the effects of attraction at higher T  for the cylinder than for either $1$d or $2$d limiting cases. Calculations using classical simulations for such gases should exhibit such an enhanced attraction effect and extend the prediction to higher densities. We anticipate that crystallization of the classical gas will occur, with an oscillatory dependence of the energy on R (due to the size-dependent commensuration energy). Such calculations are in progress in our group.

In summary, we have explored both classical and quantum particles confined to a cylindrical surface. In the case of a Coulomb interaction, which lacks a preferred distance, the energetics of binding exhibits a monotonic trend as a function of curvature; there  is an insensitivity of the cohesion to the value of the curvature, as long as it is small. In three problems involving interactions with a favored distance, quite distinct behavior was found. There is a particular range of values of R such that the binding is enhanced and this is explicable in terms of a phase space argument relevant to interactions which have well-defined potential energy minima. We believe that this distinction is generic and should be applicable to other geometries. For example, matter confined to a spherical surface (small particles or pores) should have enhanced cohesion when the diameter is$\sim \sigma$. Other related problems merit investigation. One  is the possibility of condensation of $^3$He to a liquid. The factor of $\sim 3$ enhancement of the liquid binding energy shown for cylindrical $^4$He (relative to binding in flatland) suggests that there is a range of radius over which the lighter isotopic liquid should also bind. A suggestive argument in support of that possibility is the fact that the $^3$He dimer exists over an extended range of radii: $R > \sigma  / 2 \sim  1.3$ \AA \, , according to Fig. $3$ ($\eta = 0.24$ for $^3$He). For bose systems in $1$d and $2$d, the dimer threshold coincides with the binding threshold for the many-body bound state liquid. While we do not know the corresponding criterion for  self-binding in cylinderland; it is possible that the cylindrical liquid $^3$He also exists. \footnote{
Note that out-of-plane motion will enhance the likelihood of self-binding, as was found by Whitlock $et \, al.$\cite{whi} and Gordillo  $et\, al.$\cite{gor}  for $^4$He in $2$d and $1$d and by Brami  $et\, al.$\cite{bra} for $^3$He in $2$d. The rms displacements are $0.25$ to $0.4$\, \AA \, in these cases.}
 This would be a remarkable system to investigate since it exemplifies a novel Luttinger liquid. A primary application of these results is to carbon nanotubes, whose  radius can be as small as $3$ \, \AA \, . Indeed, the so-called cylindrical shell phase of He and H$_2$ corresponds to adsorption at radial distance $\delta r \sim 3$ \, \AA \, inward from the carbon atoms \cite{gat}. Thus, the pronounced effects found here for these gases at R $\sim 2$ \, \AA \, ( Figs. $2$, $4$ ) correspond to cylindrical phases within nanotubes of radius $\sim 5 $ \, \AA \,. Experimental and other theoretical study of this size tube is worth pursuing.

We are grateful to Susana Hernandez, Claire Lhuillier, Jordi Boronat and L. W. Bruch for helpful comments. This research has been supported by Army Research Office, the Petroleum 
Research Fund of the American Chemical Society and the National Science 
Foundation (DMR0121146). Milen Kostov is grateful to Air Products and Chemicals, Inc.(APCI) 
for its support through APCI/PSU graduate fellowship.

\newpage
{\bf Figure captions}

{\bf Fig. $1$} :  Ground state energy of a ( $ + - $ ) Coulomb pair of charges, constrained on a cylinder as a function of a curvature parameter $a_0 / R$. The solid line is the variational  result.  The inset compares the variational and perturbation theory(PT) results. The energy is in Hartree units.

{\bf Fig. $2$} : Ground state cohesive energy (per atom) of a  $^4$He fluid  on a cylinder.

{\bf Fig. $3$} : Threshold value of the de Boer parameter for existence of a dimer.

{\bf Fig. $4$} : Ground state energy of $^4$He(circles) and $^3$He(stars) dimers on a cylinder.

{\bf Fig. $5$} : Reduced Boyle temperature T$^* = k_B$T/$\epsilon $ .

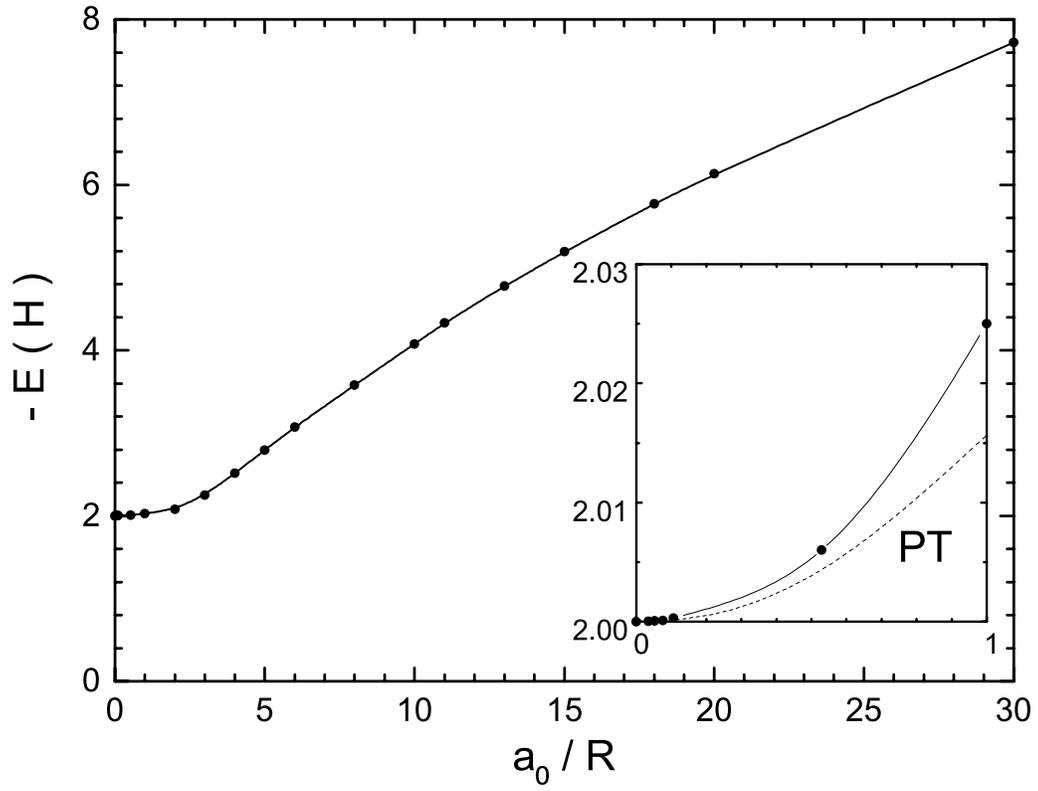
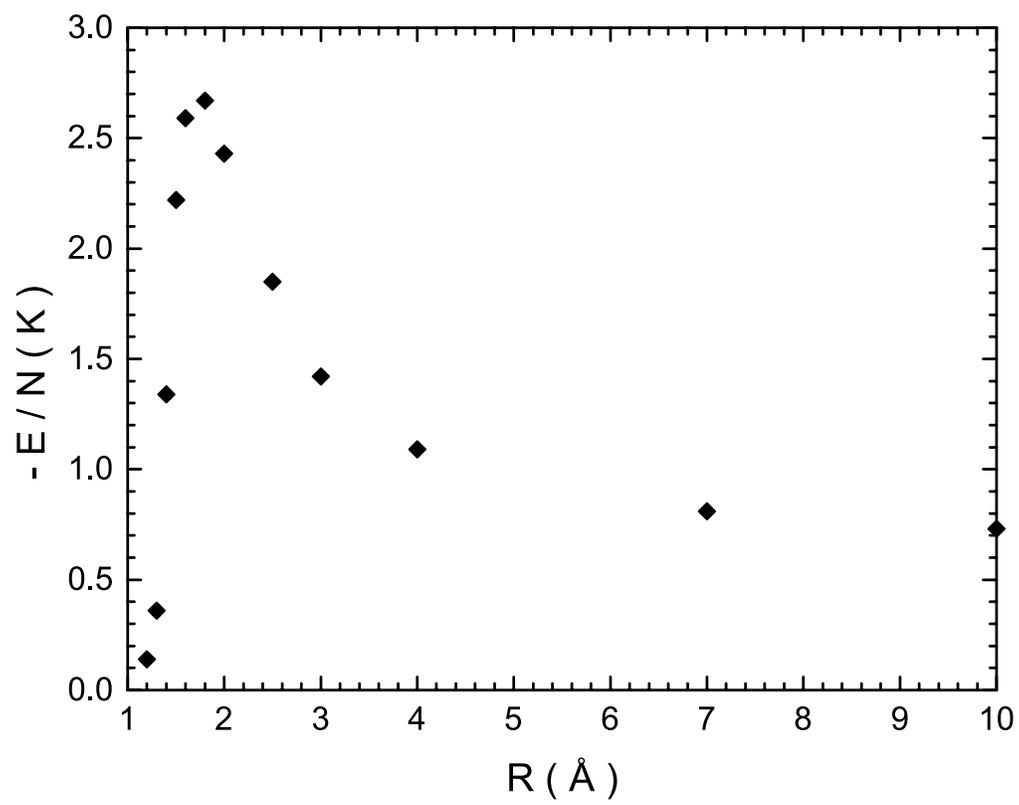
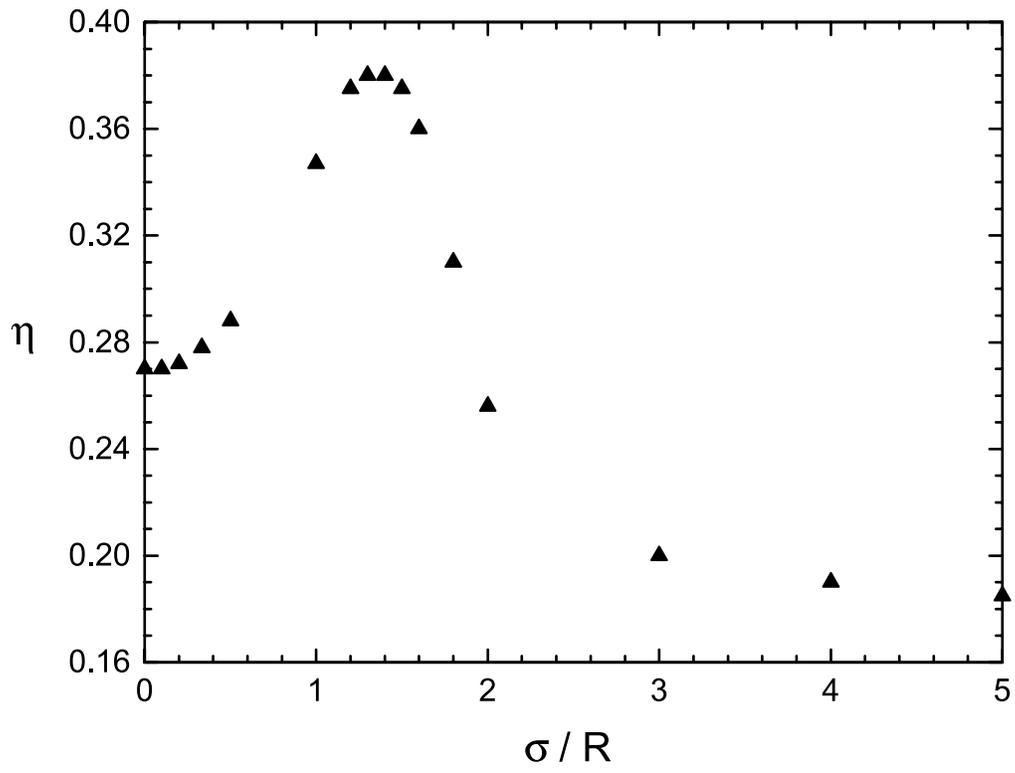
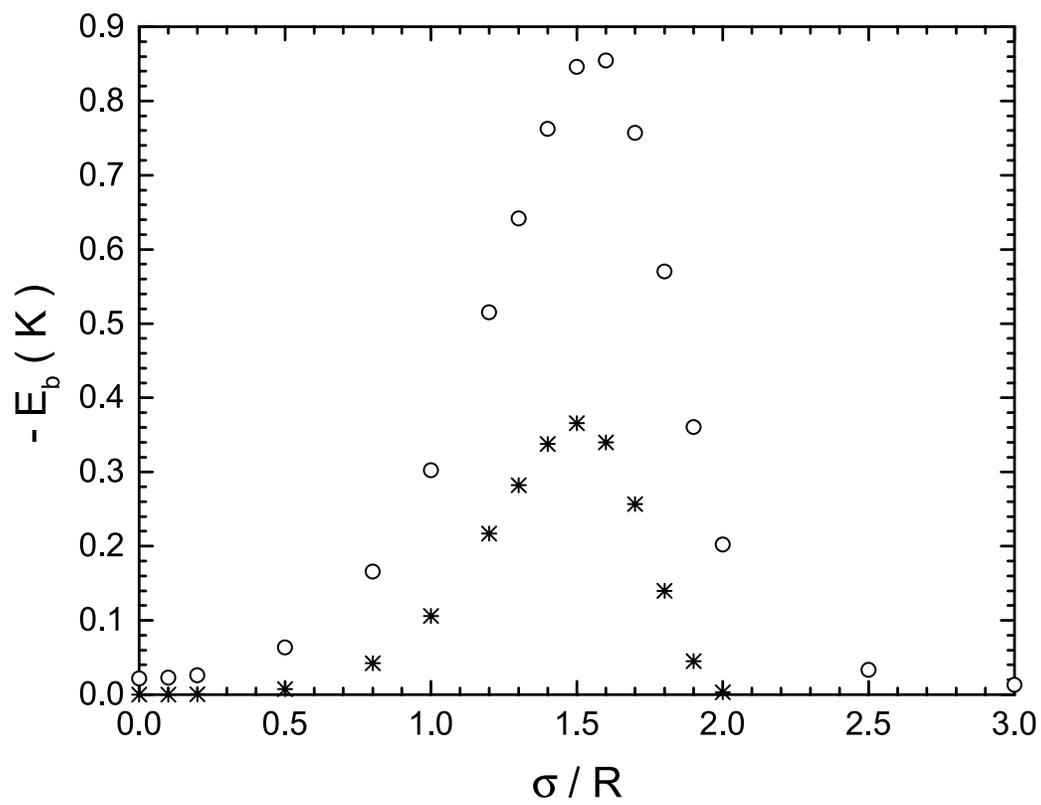
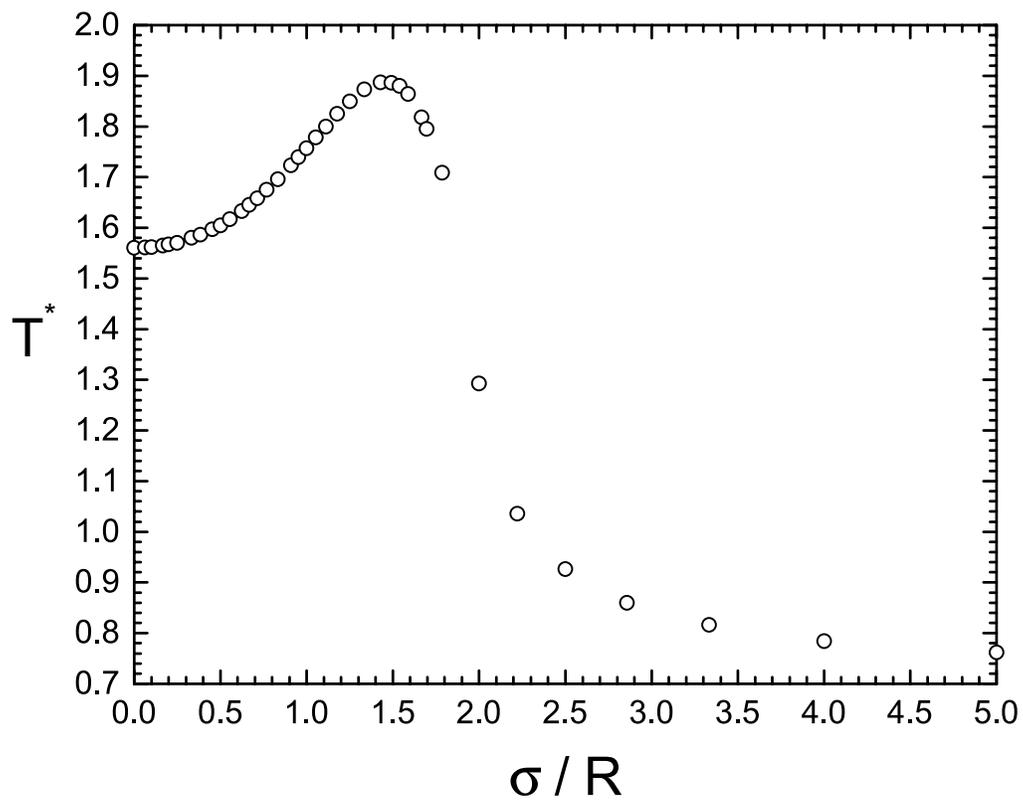
\end{document}